\documentclass[10pt,aps,prd,twocolumn,groupedaddress]{revtex4}
\usepackage{epsf}
\usepackage{dcolumn}
\newcommand{\be}{\begin{equation}}
\newcommand{\ee}{\end{equation}}
\newcommand{\bea}{\begin{eqnarray}}
\newcommand{\eea}{\end{eqnarray}}
\newcommand{\no}{\nonumber}

\begin{document}

\title{Dispersion force for materials relevant for micro and nanodevices fabrication}

\author{A Gusso}
\email{andre_gusso@yahoo.com}
\affiliation{Departamento de Ciências Exatas e Tecnol\'ogicas, Universidade Estadual de Santa Cruz\\ CEP 45662-000, Ilh\'{e}us-BA, Brazil }

\author{G J Delben}
\affiliation{ Departamento de F\'{\i}sica, Universidade Federal do Paran\'a,\\
C.P. 19044, CEP 81531-990, Curitiba-PR, Brazil}

\begin{abstract}
The dispersion (van der Waals and Casimir) force between two semi-spaces are calculated using the Lifshitz theory for different materials relevant for micro and nanodevices fabrication, namely, gold, silicon, gallium arsenide, diamond and two types of diamond-like carbon (DLC), silicon carbide, silicon nitride and silicon dioxide. The calculations were performed using recent experimental optical data available in the literature, usually ranging from the far infrared up to the extreme ultraviolet bands of the electromagnetic spectrum. The results are presented in the form of a correction factor to the Casimir force predicted between perfect conductors, for the separation between the semi-spaces varying from 1 nm up to 1 $\mu$m. The relative importance of the contributions to the dispersion force of the optical properties in different spectral ranges is analyzed. The role of the temperature for semiconductors and insulators is also addressed. The results are meant to be useful for the estimation of the impact of the Casimir and van der Waals forces on the operational parameters of micro and nanodevices.
\end{abstract}

\maketitle

%%%%%%%%%%%%%%%%%%%%%%%%%%%%%%%%%%%
\section{Introduction}
%%%%%%%%%%%%%%%%%%%%%%%%%%%%%%%%%%%

In recent years, as the size of microelectromechanical systems (MEMS) decreased, the distance between moving parts decreased correspondingly entering in the range of hundreds or even tens of nanometres.  At such separations, the relatively large moving parts of MEMS experience an attractive force due to quantum fluctuations of the electromagnetic vacuum that can be relevant for the determination of their operational parameters \cite{Chan01, GussoSNA}. Borrowing a terminology frequently used in the realm of quantum field theory  this attractive  force is commonly referred to as Casimir force \cite{review_Bordag}. Entering in the realm of nanoelectromechanical systems (NEMS),  shorter separations between moving parts are found which lie in the range of a few nanometres.  In such case, it has become usual to consider the effects of the well known van der Waals force \cite{Aluru02}. However, since the work of Lifshitz \cite{Lifshitz} the  van der  Waals and Casimir forces can be treated in a unified form as resulting from the electromagnetic quantum vacuum fluctuations existing between two separated dielectric bodies. The van der Waals(Casimir) force arises as a limiting case for short(large) separations between the bodies. Because in our analysis we are going to cover separations ranging from 1 nm up to 1 $\mu$m, we cover both limits. For that reason, and to avoid this some times confusing terminology, which includes the concept of retarded van der Waals force,  we refer to the force here analyzed simply as dispersion force, a unifying terminology also found in the literature.

The role that dispersion forces may play in the determination of the operational parameters of MEMS and NEMS has been analyzed theoretically in several instances \cite{GussoSNA, Aluru02, Maclay} and determined experimentally in, for example, references \cite{Chan01, Espinosa06, BuksRoukes}. Different methods and approximations were used in those analysis to determine the resulting dispersion force: pairwise summation of  the interatomic Lennard-Jones potential \cite{Aluru02}, a rough distance independent correction factor to Casimir force between perfectly conduction plates \cite{Maclay} or the Lifshitz theory using actual optical data \cite{Chan01, GussoSNA}  for the determination of the force between silicon plates, or  simple estimates based on the results for highly idealized cases \cite{BuksRoukes}. In fact, because the dispersion forces are highly dependent on the geometry of the interacting bodies and on their frequency dependent dielectric properties \cite{review_Bordag, Hunter, Butt} there is no simple method of calculation that applies to every situation given reliable results.

Nevertheless, for simple geometries, involving large surfaces with  small deviations from the plane parallel geometry, a powerful, reliable and relatively simple approach is to  take into account the dielectric properties of the interacting bodies through the Lifshitz theory for the pressure resulting between two semi-spaces, and the geometry through Derjaguin approximation \cite{review_Bordag, Hunter, Butt} as explained in the next section. However, results based on the Lifshitz theory \cite{Lifshitz} are available in the literature solely for three materials of relevance for micro and nanofabrication: gold and aluminium, for distances varying from less than 1 nm up to 1 $\mu$m \cite{AuAl}, and single crystal silicon, for distances varying from  1 nm up to 1 $\mu$m \cite{GussoSNA}. In the present article we expand this list including results for amorphous silicon, crystalline gallium arsenide, crystalline diamond, two types of diamond-like carbon, cubic silicon carbide, amorphous silicon nitride, and amorphous silicon dioxide, in the range 1 nm -- 1 $\mu$m.

The article is organized as follows. In section~\ref{Lifshitz} we briefly summarize how Lifshitz theory is used in order to calculate the pressure between two dielectric semi-spaces, and how this result can be used to approximate the force between dielectric non-planar bodies. In section~\ref{results} the results for the different materials we analyze are presented. We conclude in section~\ref{conclusion}.

%%%%%%%%%%%%%%%%%%%%%%%%%%%%%%%%%%%
\section{Lifshitz theory and Dejarguin approximation}
\label{Lifshitz}
%%%%%%%%%%%%%%%%%%%%%%%%%%%%%%%%%%%

According to the Lifshitz theory the dispersion force between two semi-spaces made of the same material and separated by a vacuum, is a function of the frequency dependent complex dielectric function  of the material, $\epsilon(\omega)$. The resulting pressure at a separation $d$ and at temperature $T$ can be written as \cite{Lifshitz}
\bea 
&&p^{T}(d) = - \frac{k_B T}{\pi^2 c^3} \times \no \\
  &&{\sum_{l=0}^{\infty}}^\prime \xi^3_l 
\int^\infty_1 p^2 dp 
\biggl\{\left[ \left( \frac{K  + \epsilon(i \xi_l) p}{K -
\epsilon(i \xi) p} \right)^2 e^{2(\xi_l/c)pd} -1 \right]^{-1} \no \\
 &&+ \left[ \left( \frac{K  +  p}{K  - p} \right)^2 e^{2(\xi_l/c)pd} -1
\right]^{-1} \biggr\} \, ,
\label{forceT}
\eea
were $k_B$ denotes the Boltzmann constant, $c$ the speed of light, $K=K(i \xi) = \sqrt{p^2 -1 + \epsilon(i \xi)}$, $\xi_l = 2 \pi k_B T l/\hbar$, and the prime near summation sign means that the zeroth term is taken with the coefficient 1/2. The sum in equation (\ref{forceT}) \footnote{We actually use a modified version of this expression suitable for numerical calculations. See section 5.4.2 in reference~\cite{review_Bordag}.} converges rapidly for temperatures above a few tens of Kelvin, however, for lower temperatures  the number of terms to be summed become too large and the use of the zero temperature approximation 
becomes computationally more appealing. When $k_B T \ll \hbar c/d$ we can replace the sum by an integral and the pressure becomes,
\bea
&&p^0(d) = - \frac{\hbar}{2 \pi^2 c^3} \times \no \\
 &&\int^\infty_1 p^2 dp 
\int^\infty_0 
\xi^2 d\xi \biggl\{\left[ \left( \frac{K + \epsilon(i \xi) p}{K - 
\epsilon(i \xi) p} \right)^2 e^{2(\xi/c)pd} -1 \right]^{-1} \no \\ 
 &&+ \left[ \left( \frac{K +  p}{K - p} \right)^2 e^{2(\xi/c)pd} -1 
\right]^{-1} \biggr\} \, .
\label{force0K}
\eea
At zero temperature and in the limit of perfectly conducting plates ($\epsilon \rightarrow \infty$) equation (\ref{force0K}) recovers the now classical result for the Casimir force acting between parallel plates
\be
p^C(d) = -\frac{\pi^2}{240}\frac{\hbar c}{d^4}  \, ,
\label{pC}
\ee
while for dielectrics separated by short distances the pressure results to be
\be
p^{vdW}(d) = - \frac{A_H}{6 \pi d^3},
\label{pvdW}
\ee
therefore recovering the van der Waals force ($A_H$ denotes the Hamaker constant).

Equations (\ref{forceT}) and (\ref{force0K}) depend on actual optical data through the complex dielectric function $\epsilon(\omega)= \epsilon_1(\omega)+i \epsilon_2(\omega)$.  More specifically we calculate the dielectric permittivity along the imaginary axes appearing in both  equations, $\epsilon(i \xi)$,   from $\epsilon(\omega)$ with the help of a Kramers-Kronig relation
\be
\epsilon(i \xi) = 1 + \frac{2}{\pi} \int^\infty_0 \frac{\epsilon_2(\omega)}{\omega \left(1 
+ \frac{\xi^2}{\omega^2} \right)} {\rm d}\omega \, .
\label{eps}
\ee

In order to accurately calculate the pressure over a wide range of the separation $d$, it is necessary the knowledge of $\epsilon_2(\omega)$ in a correspondingly wide range of frequencies, in such a way that the result of the approximate numerical integration performed to obtain $\epsilon(i \xi)$ be as close as possible from  the exact result.  In order to fulfill this requirement we considered the optical properties of the different materials in a range of frequencies ranging from infrared (IR) up to extreme ultraviolet (EUV) or x-ray. For all materials we consider, there is data available covering the interband transition region, in the visible and ultraviolet (UV) regions of the electromagnetic spectrum. This is the spectral region that gives the more important contribution to the dispersion force in the dielectric materials we consider. However, for large separations the contributions from IR region have to be considered while at the shortest distances we consider contributions from EUV have to be taken into account, as  is going to be evidenced by our analysis. 

For some materials we analyze there is tabulated optical data covering the entire relevant spectral region which can be interpolated an numerically integrated. However, for silicon nitride and silicon dioxide we have to perform an approximation  resorting to the Ninham-Parsegian representation \cite{Hunter} of $\epsilon(i \xi)$ in the IR region. In this representation $\epsilon(i \xi)$ is calculated  based solely on the absorption strength ($C_i$) and relaxation frequency ($\omega_i$) of the most relevant absorption peaks (in any spectral region, in principle),
\begin{equation}
\epsilon(i \xi) = 1 + \sum_{i=1}^{N} \frac{C_i}{1+(\xi/\omega_i)^2},
\label{N-P}
\end{equation}
with the second term approximating the integral in equation (\ref{eps}). Therefore, in calculating $\epsilon(i \xi)$ for silicon nitride and silicon dioxide we add the contribution of each absorption peak in the IR region.

For gallium arsenide, silicon carbide, silicon nitride, and DLC, the optical data is limited to relatively low energies situated in the UV region of the spectrum. Intending to improve the results for the shortest distances we consider we extend the data set by inclusion of the data on the index of refraction  $n$ and extinction coefficient $k$ provided by the Center of X-Ray Optics at the Lawrence Berkeley National Laboratory \cite{x-rayweb}. This data results from the compilation of both experimental and theoretical data for elements with $Z=1-92$ on the atomic scattering factors \cite{x-ray}. The resulting $n$ and $k$ (we remind the reader that $\epsilon_2 = 2 n k$) are expected to be reliable for energies above a few tens of electronvolts since at such energies the effects of atomic bounds in the solid state system, not taken into account by the atomic scattering factors, become sufficiently small. We checked this expectation by comparison with experimental data for $n$ and $k$ for several solids.  The disagreement varied from less than 5\% to crystalline silicon to more than 30\% for Germanium, however, on the average  the agreement was within 20\% for energies above a few tens of electronvolts. We comment on the consequences of using such x-ray data in section~\ref{conclusion}.

The results for the pressure between semi-infinite parallel planes can be used for the approximate calculation of the pressure between surfaces of different geometries. Whenever the  relative inclination or the curvature of the surfaces are sufficiently small and the geometries are sufficiently simple the Derjaguin approximation can be used \cite{Hunter, Butt}. This approximation permits that the results presented in the next section be used for the calculation of the dispersion forces for a large class of MEMS and NEMS made from moving parts other than parallel plates.

%%%%%%%%%%%%%%%%%%%%%%%%%%%%%%%%%%%
\section{Results}
\label{results}
%%%%%%%%%%%%%%%%%%%%%%%%%%%%%%%%%%%

In this section we present the results for the dispersion force. However, instead of the resulting pressure we present the finite conductivity correction factor, $\eta(d)$ which is a measure of the influence of the finite conductivity on the Casimir force. This factor is defined as the ratio between the pressure given by the Lifshitz theory and the Casimir pressure between perfect conductors,  equation (\ref{pC}). It evidences more clearly the effects of the different optical properties for the different materials we analyze.

For most materials we present $\eta(d)$ for different temperatures: 0 K, 300 K and, 700 K.  The chosen temperatures are representative of cryogenic temperatures, ambient temperature and the high temperatures that may be found inside integrated circuits, respectively. When only one temperature is considered we take $T =0$ K. Our results span the separation range 1 nm - 1$\mu$m. We do not consider larger distances because the dispersion forces become too small to be relevant in any practical MEMS or NEMS. On the other hand, the smallest separations we consider  are already found in sophisticated NEMS like the nanoresonator presented in \cite{Blade}, were a gap of only 20 nm exist between the moving parts. 

Besides the curve for $\eta(d)$, for the most relevant cases  we also provide a least-squares polynomial fit to the data points that can be readily used. The fitting polynomial is of the form $\eta(x) = a_0 + \sum_{n=1}^{6,7} a_n x^n$, were a new variable is used instead of $d$, namely, $x = \log_{10}(1/d)$. The fitting for all cases  are accurate within $\sim 1$\% for $d > 3 $nm, but present larger discrepancies for shorter distances, in some cases exceeding  4\% at 1 nm. The order of the polynomial is the smallest one required to keep the accuracy $\sim 1$\% for $d > 3$ nm, and varied from 6 to 7.

%%%%%%%%%%%%%%%%%%%%%%%%%%%%%%%%%%%%%%%%%%%%%%%%%
\subsection{Metals (Gold)}
%%%%%%%%%%%%%%%%%%%%%%%%%%%%%%%%%%%%%%%%%%%%%%%%%

Metals are widely used as a structural material for the fabrication of microsensors and microactuators, specially in RF MEMS, and were used in NEMS, for instance, on top of nanoresonators as conductive layers to provide a means of both induction and detection of motion \cite{Ekinci} or to produce nanocantilevers \cite{nanocantilever}, among many other applications.   Aluminium (Al) and gold (Au) are among the most widely used metals in micro and nanodevices, while nickel (Ni), cromium (Cr), titanium (Ti) and some metallic alloys were also used. 

Due to their high electrical conductivity, metals  are the materials that best approximate the perfect conductor boundary condition leading to the highest observable forces. For this reason they were used in all the recent experiments intended to measure  the Casimir force with the highest accuracy \cite{review_Bordag}.   The Casimir force for Al and Au was already investigated theoretically using the Lifshitz theory \cite{AuAl}. We present the results for gold (Au) as representative of the metals used for micro and nanofabrication to allow an immediate comparison to the results for the semiconductors and insulators we consider in the present work. Here we  use the same data set and further procedures as those employed in reference \cite{AuAl} obtaining the same results. The Casimir force was calculated at $T=0$ K using equation (\ref{force0K}). We do not consider higher temperatures because of the present controversy on how to specify the dielectric function for metals at zero frequency \cite{controversy}. However,  for a good conductor it is expected that the correction due to the non-zero temperature approaches that for perfect conductors, specially at large distances \cite{review_Bordag}, being only slightly greater. The temperature corrected pressure between perfect conductors reads, to leading order in $T$,
\begin{equation}
p^T(d) = p^0(d) \left[1 +\frac{1}{3}\left( \frac{T \, d}{1145}\right) \right]
\end{equation}
with the separation $d$ given in micrometres. According to this expression, at 300 K  the force increases by just 0.15\% at a separation of 1 $\mu$m. At 700 K, however, the correction is considerable at the same distance, the force increasing by 4.6\%. But it decreases very rapidly and at $d = 500$ nm the correction is of only 0.3\%. 
 
\begin{figure*} 
\begin{center}
\epsfysize=10cm
\epsfbox{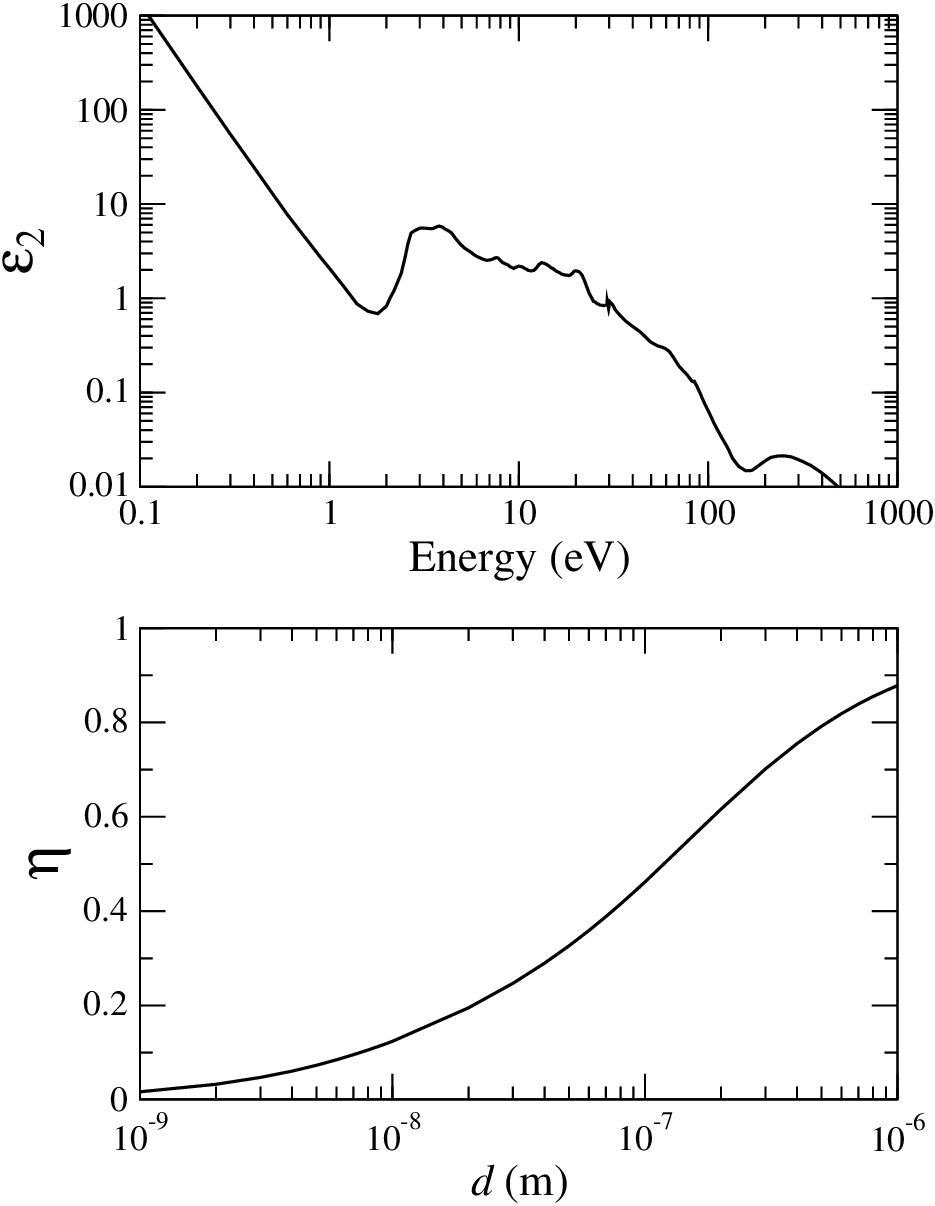}
\end{center}
\caption{\label{metal_fig}Upper panel: $\epsilon_2$ as a function of photon energy. Lower panel:  $\eta$ as a function of the separation distance $d$. Results are for gold. } 
\end{figure*}

In figure \ref{metal_fig} we present $\epsilon_2$ and $\eta$ for Au. We note that $\eta \sim 1$ at 1 $\mu$m and it decreases rapidly for shorter separations. Around 1 nm, $\eta$ changes linearly in the log-linear graph indicating that the force
varies  with $d^3$. This result is expected at short distances and characterizes the van der Waals force between two semi-spaces as seen from equation (\ref{pvdW}). The fitting function 
 for gold at 0 K is: $\eta(x) = 1.28706 \times 10^3 - 
    5.81088 \times 10^4 \, x + 1.08899 \times 10^6 \, x^2 - 
    1.08370 \times 10^7 \, x^3 + 6.03528 \times 10^7 \, x^4 - 
    1.78202 \times 10^8 \, x^5 + 2.17818 \times 10^8 \, x^6$.

%%%%%%%%%%%%%%%%%%%%%%%%%%%%%%%%%%%%%%%%%%%%%%%%%
\subsection{Silicon}
%%%%%%%%%%%%%%%%%%%%%%%%%%%%%%%%%%%%%%%%%%%%%%%%%

Silicon in both  its crystalline (c-Si) and polycrystalline forms is the most used  material in MEMS and NEMS fabrication. Crystalline silicon is widely employed is processes involving bulk micromachining, while polysilicon in those where surface micromachining is required. 
Crystalline silicon is the best understood semiconductor, and its optical properties were thoroughly investigated. High quality optical data is available from the far infrared up to x-ray energies. Here we use the tabulated data of reference \cite{Adachi} for $\epsilon_2$ for energies ranging from 0.85 eV up to 3750 eV. In the IR region the interaction of  ligth with the phonons is rather weak because silicon is a homopolar crystal and there are no active optical phonons. For this reason   $\epsilon_2$ below 0.85 eV can be  neglected \cite{Adachi}. 

As for any polycrystallyne material, the precise optical properties of polysilicon are highly dependent on the sample properties; different deposition methods and post deposition  treatments result in  different optical properties \cite{PolySi}, and the optical constants for different samples are available only for the visible spectral region. However, the  available data on $n$ and $k$ indicates that the optical properties of any polysilicon sample are going to be  between those of  c-Si and that for amorphous silicon (a-Si) \cite{PolySi}. Therefore the curves for $\epsilon_2$ and $\eta$ for c-Si and a-Si are the two limiting cases for polysilicon. For that reason and due to the fact that optical data for a-Si is available in a wide energy range,  we are going to present $\eta$ for a-Si instead of specific polysilicon samples. We use the tabulated data for a-Si of reference \cite{Adachi} ranging from 0.7 eV up to 48 eV. The contributions from the IR region can be safely neglected  as for c-Si and we do not extend $\epsilon_2$ for a-Si into the EUV region because our main interest is  simply to have a reasonable bound for $\eta$. As discussed further in section \ref{conclusion} the lack of the optical data beyond a few tens of electronvolts do not significantly affect the results for silicon for separations larger than  a few nanometres.

\begin{figure*}
\begin{center}
\epsfysize=10cm
\epsfbox{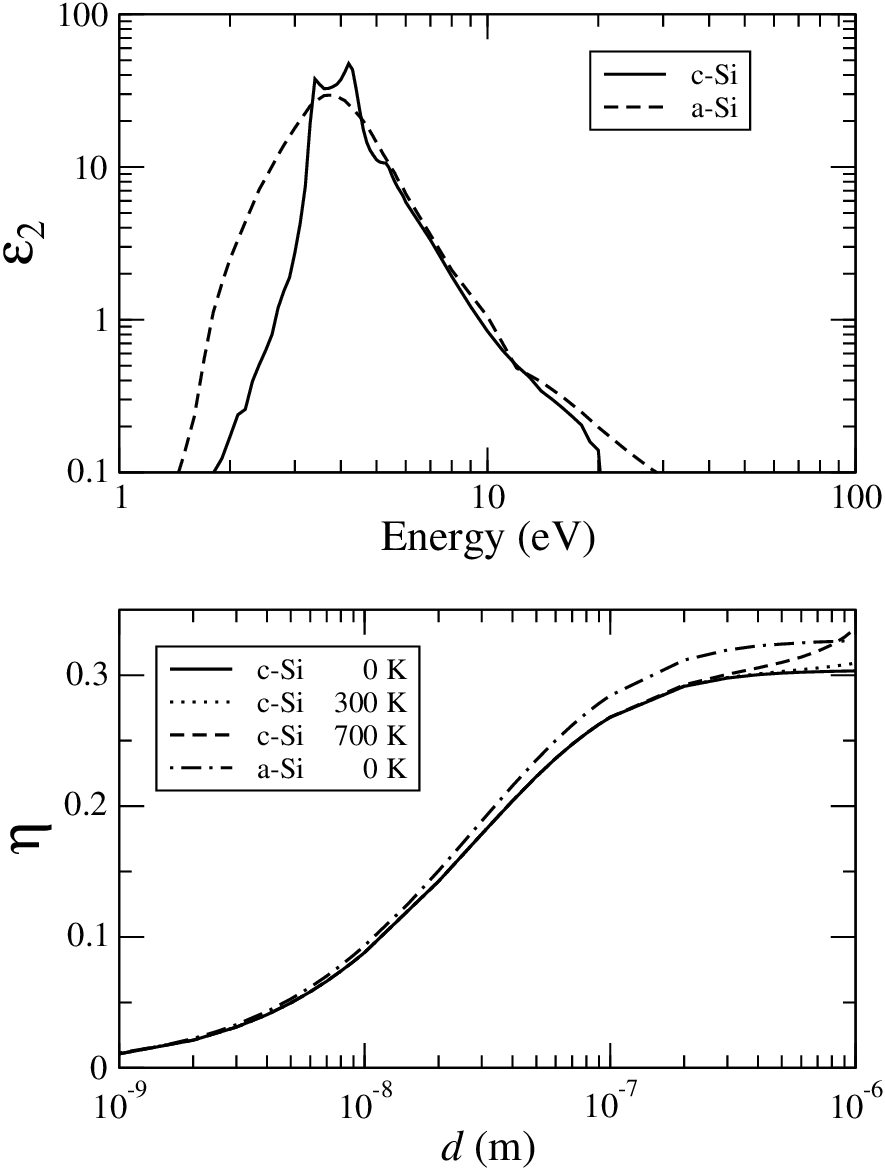}
\end{center}
\caption{\label{si_fig} Upper panel: $\epsilon_2$ as a function of photon energy. Lower panel:  $\eta$ as a function of the separation distance $d$. Results are for silicon. } 
\end{figure*}

In figure \ref{si_fig} we present $\epsilon_2$ and $\eta(d)$ for c-Si and a-Si. At $T=0$ K, $\eta$ for a-Si is greater than that for c-Si for all $d$, and we expected that $\eta$ for polysilicon samples be somewhere between the two curves. The effect of the temperature is rather small for c-Si up  $d \sim 300$ nm and becomes significant for larger separations only at $T = 700$ K. For a-Si, a similar dependence on $T$ was observed. The fitting function for c-Si at 0 K is $\eta(x) = -1.72448 \times 10^3 + 
    7.63984 \times 10^4 \, x - 1.40018 \times 10^6 \, x^2 + 
    1.35873 \times 10^7 \, x^3 - 7.36383 \times 10^7 \, x^4 + 
    2.11393 \times 10^8 \, x^5 - 2.51214 \times 10^8 \, x^6$ and for a-Si $\eta(x) = -1.76475 \times 10^3 + 
    7.81535 \times 10^4 \, x - 1.43181 \times 10^6 \, x^2 + 
    1.38893 \times 10^7 \, x^3 - 7.52482 \times 10^7 \, x^4 + 
    2.15945 \times 10^8 \, x^5 - 2.56554 \times 10^8 \, x^6$.

%%%%%%%%%%%%%%%%%%%%%%%%%%%%%%%%%%%%%%%%%%%%%%%%%
\subsection{Gallium Arsenide}
\label{GaAs}
%%%%%%%%%%%%%%%%%%%%%%%%%%%%%%%%%%%%%%%%%%%%%%%%%

Gallium Arsenide (GaAs) has been widely used on the production of MEMS and NEMS despite its, in many aspects, unattractive  mechanical properties due to its compensating attractive electrical and optical properties, availability as comercial high quality single crystal wafers, and the relatively ease control of deposition  and bulk micromachining, among other reasons \cite{Bushan}. Only the crystalline GaAs (c-GaAs) has been used in MEMS and NEMS fabrication and its amorphous form is rarely cited as a potentially useful material. For these reasons and due to the lack of optical data for amorphous GaAs \cite{Adachi} we analyze in this section solely the dispersion force for c-GaAs.

Because c-GaAs has two different atoms per unit cell and has ionic bonds, it has active optical phonons in the IR region. Therefore, contrary to silicon, we cannot simply disregard the optical properties in the IR region and we resort to the tabulated optical data collected in reference \cite{Adachi} which spans the energy range from 0.01 eV up to 15 eV, which contains one IR resonance peak. Above 15 eV, the data presented in \cite{Adachi} is only for the extinction coefficient and do not include the index of refraction. However, experimental data on $n$ and $k$  above 9.5 eV is also available in \cite{Windt}. Comparison between this data  with that collected in \cite{Adachi} evidences considerable disagreement in the overlapping energy range  corresponding to values of $\epsilon_2$ differing by more than 30\%. The differences extend over larger energies when we compare solely $k$ from both references.  Due to this incompatibility we  resort to  the x-ray data to obtain approximate results for $\epsilon_2$ in the entire energy range above 15 eV. The x-ray data is available above 30 eV and $\epsilon_2$ for the region between 15 eV and this energy was obtained by a third order interpolation including points below 15 eV and above 30 eV. We expect the error due to the use of interpolated data to be acceptably small since a relatively smooth variation of $\epsilon_2$  in this region can be expected based on the data from \cite{Windt}. 

\begin{figure*}
\begin{center}
\epsfysize=10cm 
\epsfbox{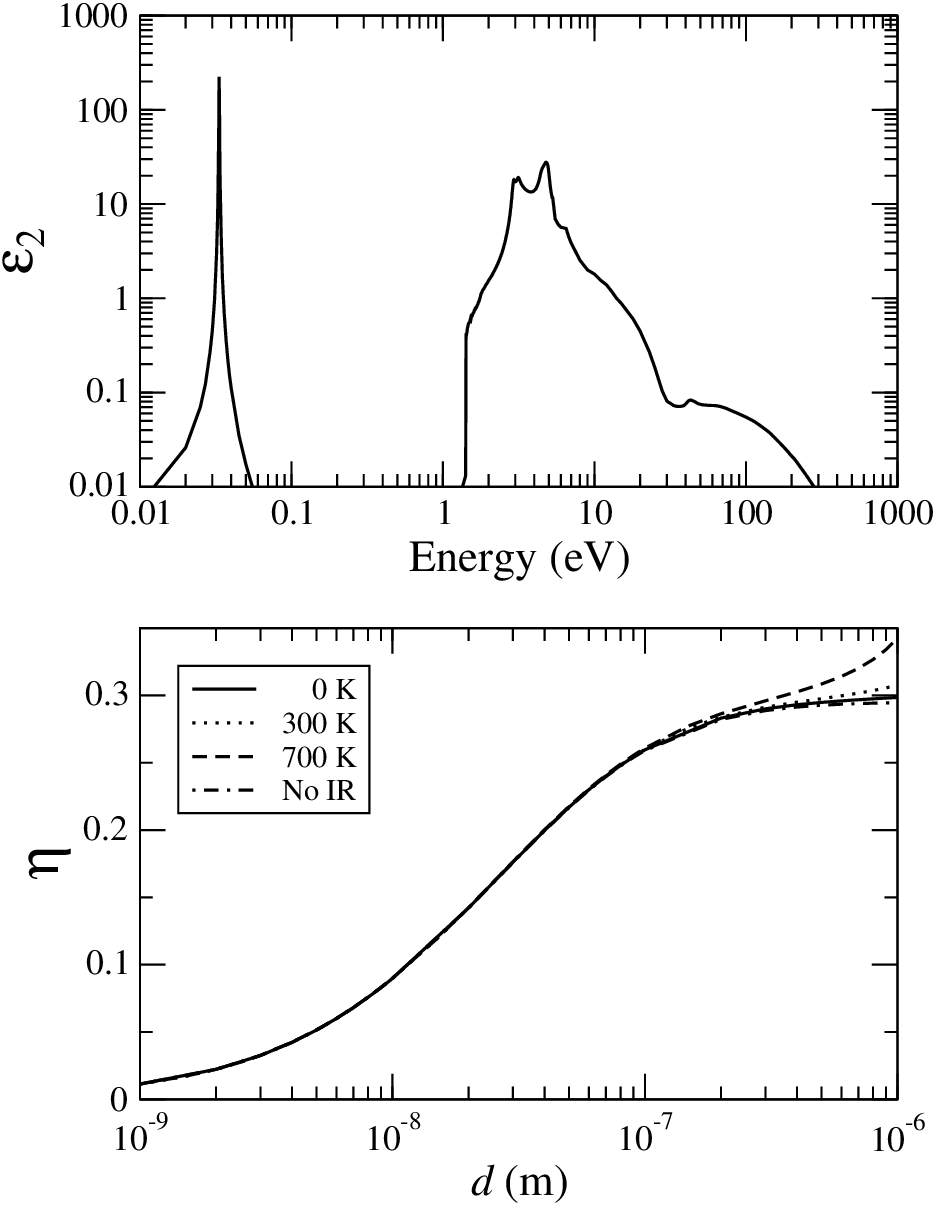}
\end{center}
\caption{\label{GaAs_fig} Upper panel: $\epsilon_2$ as a function of photon energy. Lower panel:  $\eta$ as a function of the separation distance $d$. Results are for crystalline gallium arsenide.} 
\end{figure*}

In figure \ref{GaAs_fig} we present $\epsilon_2$ and $\eta(d)$ for c-GaAs. We note how $\eta$ for c-Si and c-GaAs are quite similar with differences that rarely exceed 5\%.  The impact of the IR activity of GaAs on the dispersion force is rather small even for the largest distances we consider, leading to a dependence of $\eta$ on the temperature that is similar to that for silicon. For c-GaAs at 0 K the fitting function is $\eta(x) = -1.45323 \times 10^3 + 
    6.45764 \times 10^4 \, x - 1.18680 \times 10^6 \, x^2 + 
    1.15457 \times 10^7 \, x^3 - 6.27138 \times 10^7 \, x^4 + 
    1.80393 \times 10^8 \, x^5 - 2.14758 \times 10^8 \, x^6$.

%%%%%%%%%%%%%%%%%%%%%%%%%%%%%%%%%%%%%%%%%%%%%%%%%
\subsection{Diamond and DLC}
%%%%%%%%%%%%%%%%%%%%%%%%%%%%%%%%%%%%%%%%%%%%%%%%%

Diamond has been used on the fabrication of MEMS and NEMS in both its polycrystalline and amorphous forms (DLC). It is suited 
for devices operating in harsh enviroments due to its chemical inertness and high hardness \cite{Bushan} and for the fabrication of high frequency micro and nanomechanical resonators due to its large Young modulus, of the order of 1000 GPa, and relatively low density ($\rho \sim 3.5$ g/cm$^3$). 

Because the optical properties of polycrystalline diamond are highly sample dependent and the optical data is restricted to small spectral ranges, as for polysilicon, we do not present an analysis for this form of diamond. Instead we consider crystalline diamond (c-diamond), whose optical properties were measured from far infrared up to x-ray energies, and that can serve as a reference material. We use the tabulated data for crystalline diamond of reference \cite{Adachi} ranging from 5.5 eV up to 800 eV. $\epsilon_2$ below 5.5 eV can be safely neglected because diamond is highly transparent in the optical region and as for silicon the contributions from the IR region are small duo to the absence of active optical phonons.

DLC is a metastable form of amorphous carbon containing a significant fraction of sp$^3$ bonds \cite{DLC_review}. This terminology encompasses several forms of amorphous carbon including purely amorphous carbon (a-C) and its hydrogenated alloys (a-C:H) and the sp$^3$ bonding  rich tetrahedral amorphous carbon (ta-C) and its hydrogenated alloys (ta-C:H) \cite{DLC_review}. a-C and a-C:H have been widely used as protective coating for the disk  and read/write head in magnetic hard disks and are currently being investigated as potential materials for MEMS fabrication having its low temperature deposition as an important advantage over materials like SiC, polycrystalline and ultrananocrystalline diamond and ta-C \cite{DLC_review, a_CH}. Compared to other forms of DLC,  ta-C  is the one that best preserves the best qualities of diamond, possessing  high mechanical hardness and chemical inertness. Due to this qualities ta-C is an extremely interesting prospective material for MEMS due to its superior wear-resistant qualities, resistance to stiction ({\it i.e.} a combination of stickiness and friction) and potential as a biocompatible material that could be used inside the human body for medical purposes without generating an allergic reaction \cite{uses_ta-C}. 

The optical properties of the DLC can vary widely depending on the fraction of sp$^3$ bonds and hydrogen content. We consider two representative DLC samples, one for ta-C with a large sp$^3$ content of 80\% and another for a-C:H with a relatively low hydrogen concentration of 25\%. The optical data for $\epsilon_2$ for the ta-C sample were extracted from reference \cite{tac_data} for the energy range 1.7-40 eV. For energies above 40 eV we used x-ray data calculated considering a sample made from pure carbon and having the same density as measured in \cite{tac_data} $\rho = 3.0$ g/cm$^3$. Because there is a  mismatch between the experimental and x-ray data in the range 30-40 eV of the order of 15\% we actually use the x-ray data for energies above 50 eV softening the mismatch over the 10 eV gap between 40 and 50 eV using numerical interpolation. The optical data for a-C:H were extracted from reference \cite{ach_data} and spanned the energy range between 0.7 eV up to 30 eV. As for ta-C we observed a considerable mismatch at 30 eV between experimental and x-ray data (the x-ray data was calculated for a sample having four carbon atoms for one hydrogen and the density obtained experimentally of $\rho =$ 1.75 g/cm$^3$.) and used a 10 eV gap in order to soften the mismatch. Therefore, x-ray data was used above 40 eV. With regard to  the IR activity of ta-C and a-C:H, it was experimentally verified \cite{IR_DLC} that both forms of DLC have low absportion in the IR region, and therefore $\epsilon_2$ in the IR can be neglected.

\begin{figure*}
\begin{center}
\epsfysize=10cm
\epsfbox{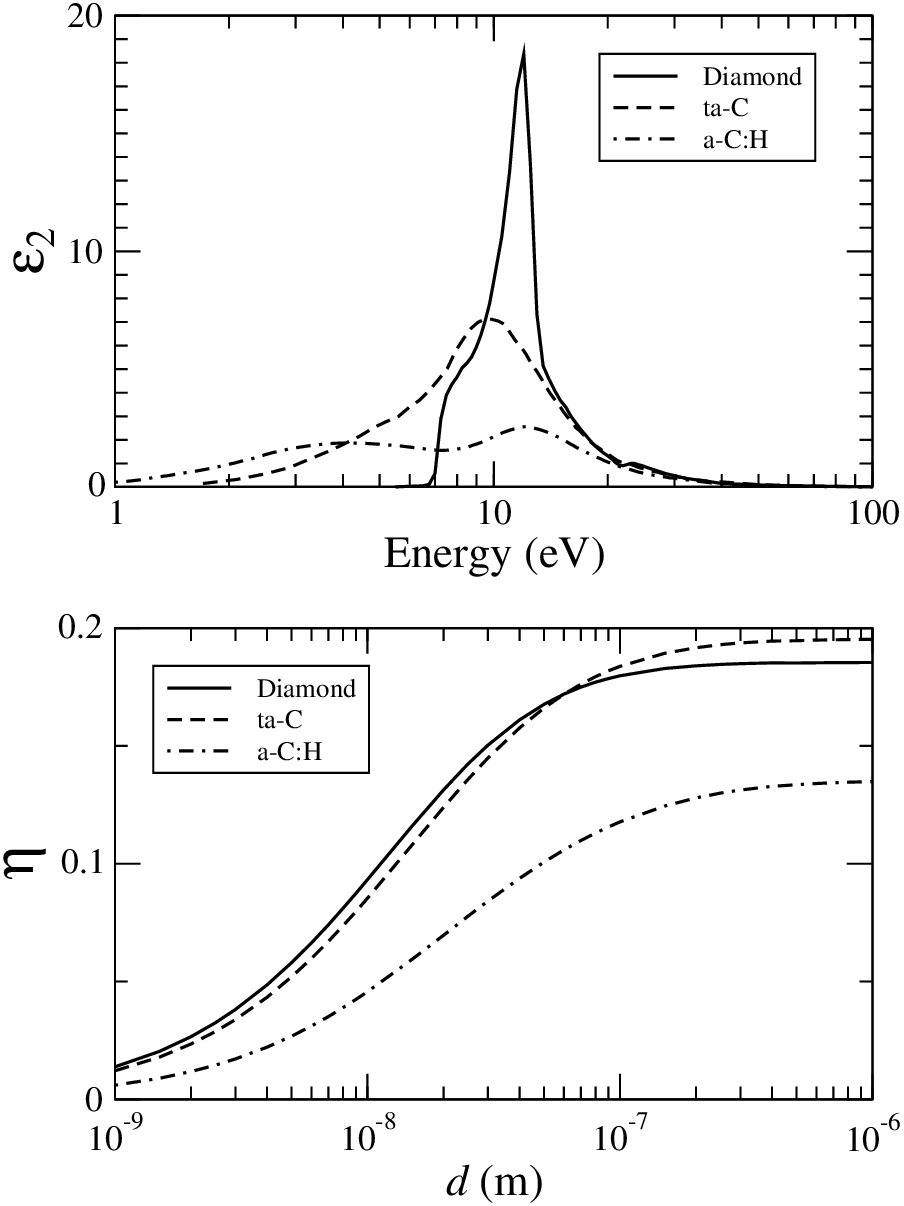}
\end{center}
\caption{\label{diamond_fig} Upper panel: $\epsilon_2$ as a function of photon energy. Lower panel:  $\eta$ as a function of the separation distance $d$. Results are for diamond.} 
\end{figure*}

In figure \ref{diamond_fig} we present $\epsilon_2$ and $\eta(d)$ for c-diamond and the two forms of DLC.  
The resulting correction factors $\eta$ present an interesting behavior. At short distances, below  approximately 100 nm the Casimir force for c-diamond is slightly larger than that for ta-C, however for separations above 100 nm an inversion occurs.  This inversion can be easily understood. The force at short distances is influenced mostly by the strength of the resonances in the UV region which are clearly weaker for  ta-C as compared to c-diamond. At larger distances the force is influenced by lower frequency modes for which $\epsilon_2$ is larger in the case of ta-C . It is interesting to note also the decrease in the dispersion force for a-C:H  relative to c-diamond and ta-C, making a-C:H an interesting choice to minimize the Casimir force. 

The fitting function for c-diamond at 0 K is the 7th order polynomial $\eta(x) = -6.88550 \times 10^3 + 
    3.53738 \times 10^5 \, x - 7.74634 \times 10^6 \, x^2 + 
    9.37393 \times 10^7 \, x^3 - 6.77091 \times 10^8 \, x^4 + 
    2.91985 \times 10^9 \, x^5 - 6.96200 \times 10^9 \, x^6 +
    7.08214 \times 10^9 \, x^7$.

%%%%%%%%%%%%%%%%%%%%%%%%%%%%%%%%%%%%%%%%%%%%%%%%%
\subsection{Silicon Carbide}
%%%%%%%%%%%%%%%%%%%%%%%%%%%%%%%%%%%%%%%%%%%%%%%%%

Silicon carbide (SiC) is a material with very attractive properties for MEMS and NEMS applications similar to those of diamond. Its mechanical  strength,  high  thermal  conductivity,  ability  to  operate  at high  temperatures  and extreme  chemical  inertness, make  SiC   attractive for  MEMS and NEMS applications in harsh enviroments,  both  as  structural  material  and  as  coating  layer, and for high frequency micro and nanomechanical resonators. SiC  has been used on the fabrication of MEMS and NEMS in both its  crystalline  and polycrystalline (poly-SiC) forms \cite{Bushan} and the potential usage of its amorphous form has been investigated \cite{amorphousSiC}. In its crystalline form, SiC is a polymorphic material that exist in cubic, hexagonal, and rhombohedral polytypes. However, only for the cubic polytype (3C-SiC) there exist adequately measured optical data \cite{Adachi} in a sufficiently wide range of energies ranging from 0.01 eV up to 20 eV \cite{Adachi, SiCdata}. Therefore, due to the lack of data for polycrystalline and amorphous SiC, as well as for the non-cubic polytypes  we restrict our analysis and consider the Casimir force for crystalline 3C-SiC. 

Because SiC is a heteropolar semiconductor it is active in the IR region. 3C-SiC presents a strong IR resonance around 0.0987 eV due to an active transversal optical phonon. We use the data collected in \cite{Adachi} between 0.01 eV up to 0.4 eV which covers the IR region containing the resonance just mentioned. For energies above 0.4 eV and up to approximatelly 20 eV we use the data on $\epsilon_2$ of reference \cite{SiCdata}. Above 30 eV we resort to x-ray data, and use an interpolating function in the 20-30 eV range.

\begin{figure*}
\begin{center}
\epsfysize=10cm
\epsfbox{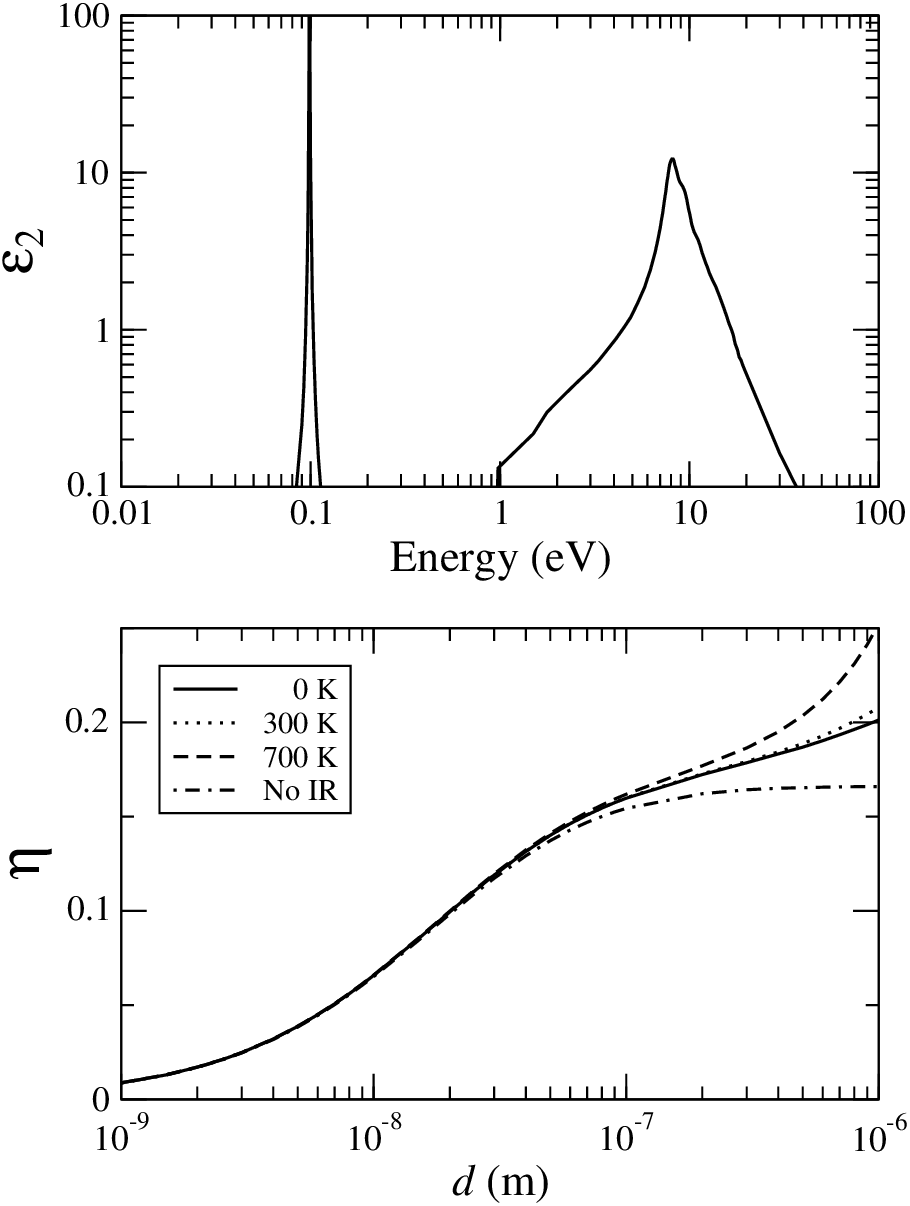}
\end{center}
\caption{\label{SiC_fig} Upper panel: $\epsilon_2$ as a function of photon energy. Lower panel:  $\eta$ as a function of the separation distance $d$. Results are for silicon carbide.} 
\end{figure*}

In figure \ref{SiC_fig} we present $\epsilon_2$ and $\eta(d)$ for 3C-SiC.
In order to determine the impact of the IR activity of the SiC on $\eta(d)$ we present in figure \ref{SiC_fig} $\eta$ calculated at 0 K without the contributions from the IR region (indicated by the label ``No IR" in the figure). The result is considerably smaller than that for the full spectrum at 0 K for distances above approximately 100 nm, demonstrating the importance of the IR activity for this material. More specifically, the relative increase due to the IR activity on the force is about 21\% at a separation of 1 $\mu$m. For non-zero temperature we observe a larger relative increase of $\eta$ as compared to the materials analyzed previously, specially for distances near 1 $\mu$m were it increases by 25\%  at 700 K  compared to 0 K. This larger increase can be attributed to the contribution of the IR region. The different consequences on $\eta$ of the IR activity for GaAs and SiC can be attributed to the much stronger IR resonance peak of the later (by about a factor of six) and to its occurrence at a higher energy around 0.1 eV, compared to 0.033 eV for GaAs.

As for c-diamond the fitting function for 3C-SiC at 0 K is a 7th order polynomial, namely, $\eta(x) = -4.67473 \times 10^3 + 
    2.36272 \times 10^5 \, x - 5.08848 \times 10^6 \, x^2 + 
    6.05401 \times 10^7 \, x^3 - 4.29817 \times 10^8 \, x^4 + 
    1.82144 \times 10^9 \, x^5 - 4.26710 \times 10^9 \, x^6 +
    4.26431 \times 10^9 \, x^7$.

%%%%%%%%%%%%%%%%%%%%%%%%%%%%%%%%%%%%%%%%%%%%%%%%%
\subsection{Silicon Nitride}
%%%%%%%%%%%%%%%%%%%%%%%%%%%%%%%%%%%%%%%%%%%%%%%%%

Silicon nitride (SiN) is widely used in MEMS for electrical
isolation, surface passivation, etch masking, and as a mechanical material \cite{Bushan}, being in the last case also used in NEMS, for instance, on the fabrication of nanoelectromechanical oscillators \cite{Si3N4nems}. Silicon nitride is usually employed in its amorphous form and frequently it is also non-stoichiometric and may contain significant concentrations of hydrogen. Good optical data is available in a wide energy range for the amorphous stoichiometric silicon nitride (a-Si$_3$N$_4$) \cite{HOC}, and for  crystalline stoichiometric  $\beta-$Si$_3$N$_4$ \cite{Si3N4_French}. Here we focus on the results for a-Si$_3$N$_4$ because SiN is usually employed in its amorphous form in MEMS and NEMS.

For a-Si$_3$N$_4$ we use the data from \cite{HOC} which spans the energy range 4-24 eV. There are two resonances in the IR region whose effects we introduce using equation (\ref{N-P}) with $C_1 = 1.08$ and $\omega_1 = 0.9 \times 10^{14}$ rad s$^{-1}$, and  $C_2 = 2.37$ and $\omega_2 = 1.64 \times 10^{14}$ rad s$^{-1}$ \cite{Bergstrom}. For energies above 30 eV we use x-ray data calculated for stoichiometric Si$_3$N$_4$  with density $\rho = 3.2$ g cm$^{-3}$, and use cubic interpolation between 24 and 30 eV. We note that the $\epsilon_2$ generated using x-ray data has very good agreement with the experimental data for $\beta$-Si$_3$N$_4$.

\begin{figure*}
\begin{center} 
\epsfysize=10cm
\epsfbox{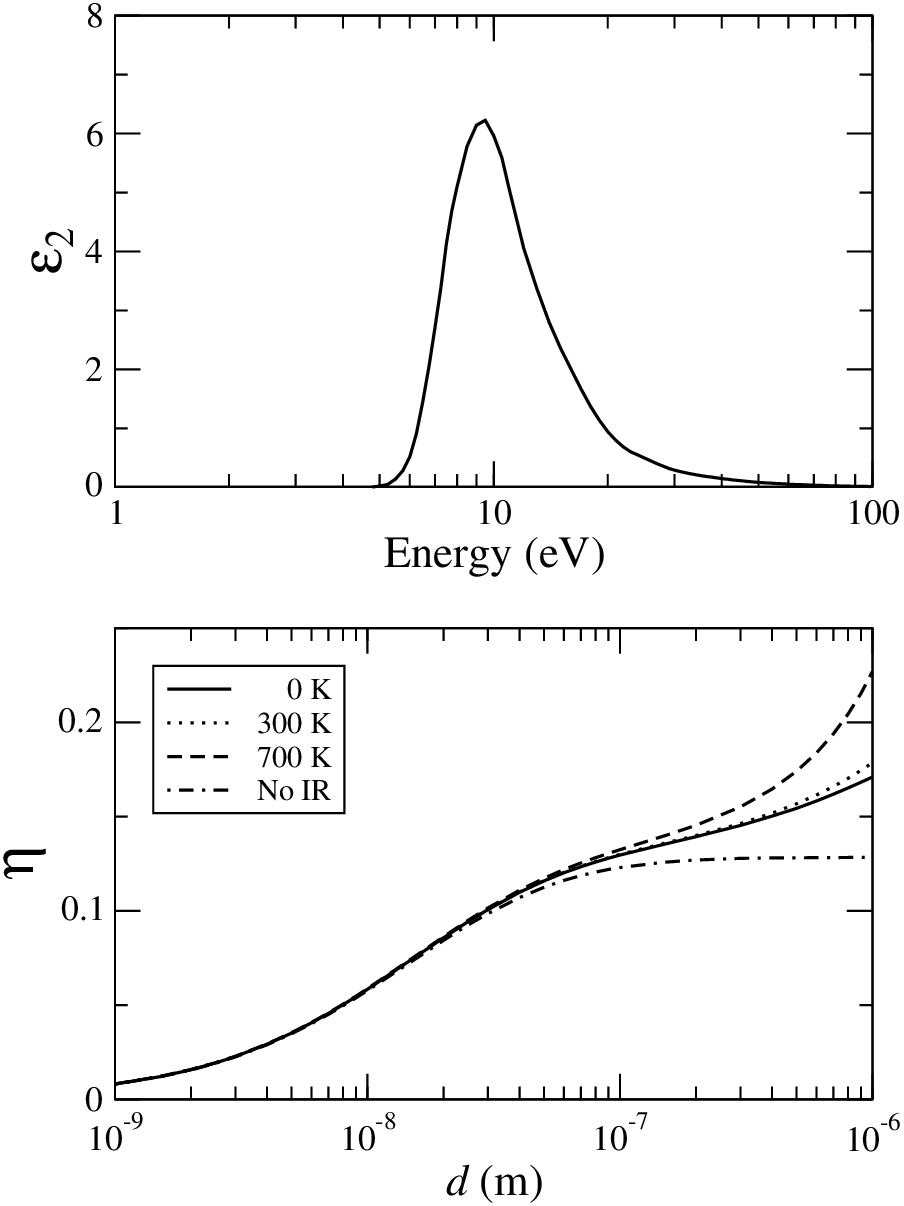}
\end{center}
\caption{\label{SiN_fig} Upper panel: $\epsilon_2$ as a function of photon energy. Lower panel:  $\eta$ as a function of the separation distance $d$. Results are for amorphous silicon nitride.} 
\end{figure*}

In figure \ref{SiN_fig} we present $\epsilon_2$ and $\eta(d)$ for a-Si$_3$N$_4$. From the results for $\eta$ it can be concluded that the resulting Casimir force is smaller than that for the previously analyzed materials. The influence of the IR activity is also noticeable, as for SiC, having a  larger relative importance on $\eta$, that corresponds to a 34\% increase on the force at 1 $\mu$m and 0 K.  The effect of temperature is also large for a-Si$_3$N$_4$, leading to a 33\% increase on the force at 1 $\mu$m and 700 K as compared to 0 K.

The fitting function for a-Si$_3$N$_4$ at 0 K is a 7th order polynomial, namely, $\eta(x) = -4.57954 \times 10^3 + 
    2.33430 \times 10^5 \, x - 5.07148 \times 10^6 \, x^2 + 
    6.08838 \times 10^7 \, x^3 - 4.36263 \times 10^8 \, x^4 + 
    1.86622 \times 10^9 \, x^5 - 4.41385 \times 10^9 \, x^6 +
    4.45362 \times 10^9 \, x^7$.

%%%%%%%%%%%%%%%%%%%%%%%%%%%%%%%%%%%%%%%%%%%%%%%%%
\subsection{Silicon Dioxide}
%%%%%%%%%%%%%%%%%%%%%%%%%%%%%%%%%%%%%%%%%%%%%%%%%

Silicon dioxide, in its amorphous form (a-SiO$_2$), is usually known for its applications in microelectronics as an electric insulator. More recently it has been widely used in the fabrication of MEMS and NEMS as a sacrificial layer material and has also found application as a structural material, for instance, for the fabrication of large suspended membranes and grating light valves \cite{GLV}. Quartz, the crystalline form of SiO$_2$, has also  been used as a structural material for the fabrication of MEMS, specially in microfluidics.

For both a-SiO$_2$ and quartz there is good optical data available in a wide range of energies and we resort to the most recent works. In \cite{Frenchsio2} the data covers the range 1.5-42 eV for both forms of SiO$_2$, while in \cite{Filatovasio2} the energy range  65-3000 eV is covered for the amorphous form. Due to the availability of recent data in a wider range for  a-SiO$_2$ and to its almost omnipresence in MEMS and NEMS we analyze the Casimir force for this form of silicon dioxide. However, due to the similarity between the optical properties of crystalline and amorphous SiO$_2$ evidenced in \cite{Frenchsio2}, our results are expected to be approximately valid for quartz.

Complementing the experimental data from  \cite{Frenchsio2, Filatovasio2} we use x-ray data between 42 eV and 65 eV, which has shown a very good agreement with both data sets between 30 eV and approximately 100 eV where data superposition exists. From the IR region we considered the existence of three resonances  whose effects we introduce using equation (\ref{N-P}) with $C_1 = 0.829$ and $\omega_1 = 0.867 \times 10^{14}$ rad s$^{-1}$,  $C_2 = 0.095$ and $\omega_2 = 1.508 \times 10^{14}$ rad s$^{-1}$, and $C_3 = 0.798$ and $\omega_3 = 2.026 \times 10^{14}$ rad s$^{-1}$  \cite{Bergstrom}.

\begin{figure*}
\begin{center} 
\epsfysize=10cm
\epsfbox{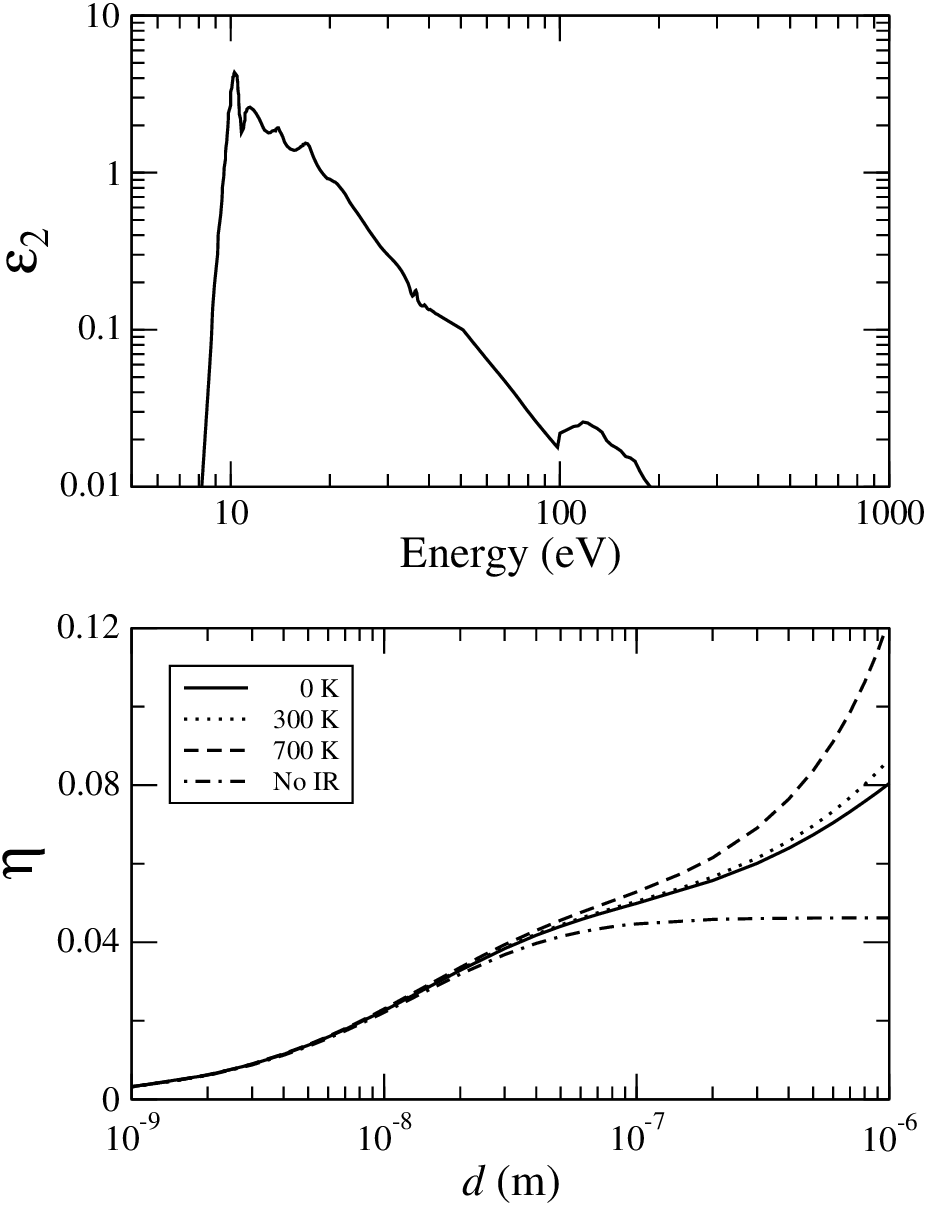}
\end{center}
\caption{\label{SiO2_fig} Upper panel: $\epsilon_2$ as a function of photon energy. Lower panel:  $\eta$ as a function of the separation distance $d$. Results are for amorphous silicon dioxide.} 
\end{figure*}

In figure \ref{SiO2_fig} we present $\epsilon_2$ and $\eta(d)$ for a-SiO$_2$. From the results for $\eta$ it can be concluded that the resulting dispersion force is considerably smaller than that for the previously analyzed materials, a result attributable to the largest band gap of about 8.3 eV \cite{Phillipssio2}. The influence of the IR activity is also the strongest amongst all the materials we analyzed increasing   the force by 74\% at 1 $\mu$m  and 0 K.  The effect of temperature is also rather large, leading to a 52\% increase on the force at 1 $\mu$m and 700 K as compared to 0 K.

The fitting function for SiO$_2$ at 0 K is a 7th order polynomial, namely, $\eta(x) = -1.84016 \times 10^3 + 
    9.41606 \times 10^4 \, x - 2.05374 \times 10^6 \, x^2 + 
    2.47524 \times 10^7 \, x^3 - 1.78058 \times 10^8 \, x^4 + 
    7.64622 \times 10^8 \, x^5 - 1.81524 \times 10^9 \, x^6 +
    1.83827 \times 10^9 \, x^7$.

%%%%%%%%%%%%%%%%%%%%%%%%%%%%%%%%%%%%%%%%%%%%%%%%%
\subsection{Mixed materials}
%%%%%%%%%%%%%%%%%%%%%%%%%%%%%%%%%%%%%%%%%%%%%%%%%

\begin{figure*}
\begin{center} 
\epsfysize=5cm
\epsfbox{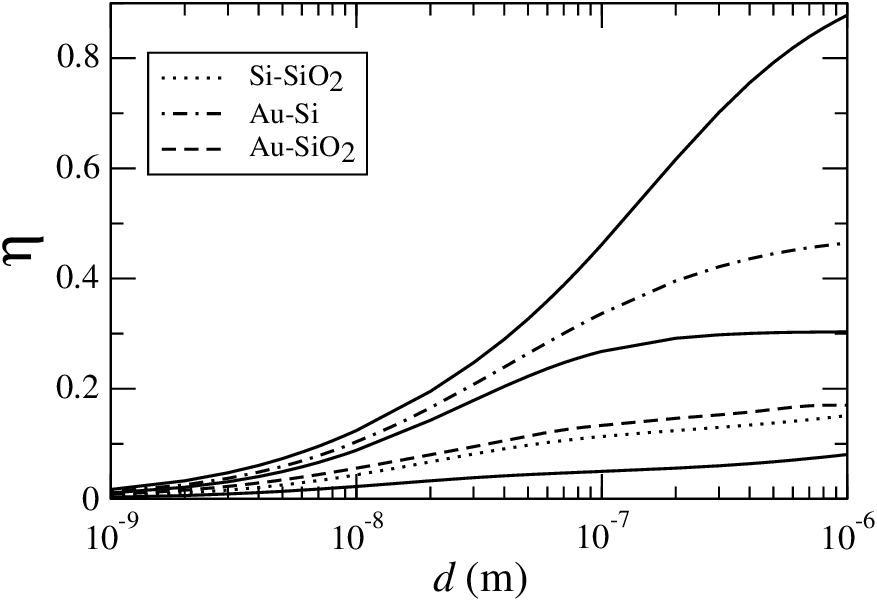}
\end{center}
\vspace{1.5cm}
\caption{\label{mixed_fig} $\eta$ as a function of the separation distance $d$. Results are for two different materials on each semi-space. The continuous lines are (downward) for Au, c-Si and a-SiO$_2$. } 
\end{figure*}

As our last result, we present $\eta$ for combinations of different materials in each semi-space. In this case we use a generalization of equation (\ref{force0K}) for the pressure between semi-spaces made from different materials found, for instance, in \cite{Milonni}.  As representative of all possible combinations we consider the following three combinations of Au, c-Si and a-SiO$_2$: Au-Si, Au-SiO$_2$ and Si-SiO$_2$. They represent the  three possible combinations  conductor-semiconductor, conductor-insulator, and semiconductor-insulator. The results for $\eta$ are presented in the figure \ref{mixed_fig}, where, as a reference, we also reproduce the results for Au, Si and SiO$_2$. It can be seen that, as could be expected, the resulting force is always between that produced for each material separately.

%%%%%%%%%%%%%%%%%%%%%%%%%%%%%%%%%%%%%%%%%%%%%%%%%
\section{Discussions and conclusion}
\label{conclusion}
%%%%%%%%%%%%%%%%%%%%%%%%%%%%%%%%%%%%%%%%%%%%%%%%%

In performing our calculations we used the most recent and complete optical data available for each material in order to derive the most accurate results. However, even the best tabulated optical data relies on results from different experiments based on different samples and methodologies covering different, and in many cases, not overlapping spectral regions that must be interpolated  or modelled theoretically \cite{Adachi, HOC}. Besides there may exist a significant disagreement between experiments, of the order of tens of percent,  as pointed in section~\ref{GaAs} for GaAs and as verified by the authors also for the case of 3C-SiC and a-SiO$_2$ when the results from references \cite{Adachi} and \cite{SiCdata} and \cite{Frenchsio2} and \cite{sio2_eels} were compared, respectively.  Therefore an analysis of the expected uncertainties in our theoretical predictions is necessary. 

The result for gold must be considered taking into account the thorough analysis performed in \cite{analysis_gold} that has shown that great uncertainties, around 5\%, must be assumed for the predicted force between gold plates.  However, the analysis performed for gold can not be extended for semiconductors and insulators.  We have firstly to rely on the analysis performed by  Aspnes and Studna \cite{AspnesStudna} in 1983 in order to state  that  for the well studied c-Si, GaAs, c-diamond and a-SiO$_2$ the uncertainty on the latest optical data for $\epsilon_2$ is about  2\%.   This is an estimate based on the analysis of the convergence of different data sets, and means that better measurements should not result in changes for the optical data larger than this uncertainty. This is the only uncertainty estimate we can perform since, as a rule, the optical data presented in the references we used are not accompained by an analysis of theoretical or experimental uncertainties. For a-Si, 3C-SiC and a-Si$_3$N$_4$ larger uncertainties on $\epsilon_2$ should be assumed because few experiments were performed to measure their optical properties. We suggest that an uncertainty of at least 5\% be assumed. In the case of DLC, the dependence of optical properties on  sample preparation make it worth to consider our results only as a good estimate to the dispersion force. 

The resulting uncertainty on $\eta$ ($\delta \eta$) due to the uncertainty on $\epsilon_2$ ($\delta \epsilon_2$) depends on the separation $d$ and material under consideration. For instance, for silicon $\delta \eta$ is always smaller than $\delta \epsilon_2$, being within 40\% to 80\% of its value. However, for 3C-SiC and a-Si$_3$N$_4$ $\delta \eta$ results to be larger than $\delta \epsilon_2$ for short distances, reaching a maximum of $\delta \eta = 5.8$\% for $\delta \epsilon_2 = 5$\% in the case of a-Si$_3$N$_4$. For larger distances $\delta \eta$ decreases to approximatelly 70\% of $\delta \epsilon_2$ for both materials. As a reference we present in table~\ref{errors} the maximum and minimum $\delta \eta$ as a function of $\delta \epsilon_2$ for the most relevant semiconductors and insulators we considered. The general trend for the uncertainty is that it decreases rapidly from its maximum at 1 nm to a value close to its minimum for $d \sim 100$ nm and is almost a constant for larger $d$ reaching a minimum at 1 $\mu$m.

\begin{table}
\caption{\label{errors} Maximum uncertainty on $\eta$ due to variations on $\epsilon_2$. Approximate values for small and large separation $d$.}
\begin{tabular}{@{}lll}
Material($\delta \epsilon_2$) & $\delta \eta$ (Small $d$)  & $\delta \eta$ (Large $d$) \\ \hline
c-S (2\%) &  1.6\%  &   0.9\%  \\
c-GaAs (2\%)  &  1.7\%  &  0.9\%  \\
c-diamond (2\%) &  2.0\%  &  1.4\%  \\
3C-SiC (5\%)  &  5.3\%  & 3.2\%  \\
a-Si$_3$N$_4$ (5\%) & 5.8\%  &  3.6\%  \\
a-SiO$_2$ (2\%) & 2.9\%   &  2.0\%  \\ \hline
\end{tabular}
\end{table}

For c-GaAs, 3C-SiC, a-Si$_3$N$_4$ and DLC another source of uncertainty results from the use of x-ray data.  As already mentioned in section~\ref{Lifshitz} comparison between the experimental and x-ray data for several materials, like aluminium, copper, silicon and germanium, evidences that x-ray data is precise within about 20\% for energies above 30 eV. The actual disagreement could be smaller, since the x-ray data tends to agree well (within 10\%) with the most recent, and presumably more precise, experimental data for materials like silicon and silicon dioxide.  We can assume conservatively that $\delta \epsilon_2$ for such energies is about 20\%.  In order to estimate how this uncertainty affects $\eta$, we firstly note that for the semiconductors and insulators we are considering $\epsilon_2$ above 30 eV amounts to at most 6\% of $\eta$ for $d \sim 1$ nm, 3\% at $d \sim 10$ nm and contributes less than 1.5\% at 100 nm. Therefore, the extra uncertainty due to the x-ray data is at most approximately 1.2\% at 1 nm, 0.6\% at 10 nm and 0.3\% at 100 nm, and can be neglected for larger separations.   It has to be noted, that this extra uncertainty results in an error that is much smaller than the error resulting from the neglect of $\epsilon_2$ above 30 eV, therefore justifying the use of the x-ray data when no experimental data is available. 

The above uncertainty analysis demonstrates that a complex relationship exists between the uncertainty in the optical data and the theoretical prediction of the Casimir force. Therefore, for precise experiments a careful analysis must be performed for each material and experimental condition. However, the results we present are useful for theoretical estimates of the impact of dispersion force on the operational parameters of MEMS and NEMS.  

We also addressed the role of temperature on the expected dispersion force since MEMS and NEMS operating in real conditions may be subject to high temperatures that alter the force. For all materials but SiO$_2$ the effect of the ambient temperature of $T=300$ K on the force is rather small. At $T=700$ K for silicon, gallium arsenide, and diamond  the effect of temperature is not so severe but exceeds the approximate result for metals by at least a factor of two for $d \sim 1 \mu$m. However, for the materials with strong IR activity SiC, Si$_3$N$_4$ and SiO$_2$ the effects of the temperature are already  noticiable at shorter distances for $T = 700$ K and are rather large for $d$ in the range of hundreds of nanometres.  At this point, it is worth to mention that no experiment performed to date has measured the dependence of the dispersion force on temperature but, according to our results, materials with strong IR activity are the most adequate for such an experiment due to the greater sensitivity of the  force on temperature.

Our analysis also evidences how  significant  can be  the contribution  for the dispersion force coming  from the IR region through the results for  SiC, Si$_3$N$_4$ and SiO$_2$.  The effects of the IR activity are noticiable already at distances in the range of a few tens of nanometres and increase the dispersion force by  tens of percent at larger separations even at 0 K. Therefore, due to its potential effects on the dispersion force,  knowledge of the IR optical behavior of a given material is mandatory when this force is to be calculated for distances ranging from tens up to hundreds of nanometres. The IR spectral region can be neglected only after a careful analysis.

When the dispersion forces are to be calculated for short distances, smaller than about 100 nm, knowledge of the optical properties in the UV region is required. We considered tipically the optical properties up to a few hundreds of electronvolts, corresponding to the EUV spectral region.  However, as expected for semiconductors and insulators, the more important contribution to the dispersion force over the entire interval considered for the separation $d$  comes from the interband transition region, usually extending from $\sim 1$ eV up to $\sim 20$ eV. 

To conclude, we note that the dispersion forces for the different materials relevant for micro and nanodevices fabrication can vary considerably. When large dispersion forces are desirable, good conductors or small band gap semiconductors should be used. However, in order to minimize the dispersion force, insulators are recommended. The results for diamond and DLC are an example on how the dispersion force can be taylored by chemical or physical changes   of a given material as a strategy to control the dispersion forces in MEMS and NEMS.

\begin{acknowledgments}
The authors were supported by the Conselho Nacional de Desenvolvimento
Cient\'{\i}fico e Tecnol\'ogico (CNPq-Brazil).
\end{acknowledgments}

\end{document}